\documentclass[floatfix,aps,prb,preprint,superscriptaddress,showpacs]{revtex4-1}
\usepackage{amsmath,amssymb,graphicx}

\begin{document}

\def\r{{\bf r}}

\title{A fully quantum mechanical calculation of the diffusivity of
  hydrogen in iron using the tight binding approximation and path
  integral theory}

\author{I. H. Katzarov}
\altaffiliation[Present address: ]{ Department of Physics, King's 
  College London, Strand, London WC2R 2LS, UK}
\affiliation{Atomistic Simulation Centre, School of Mathematics
  and Physics, Queen's University Belfast, Belfast BT7 1NN, UK}
\author{A. T. Paxton}
\affiliation{National Physical Laboratory, Teddington, Middlesex, TW11 0LW, UK}
\affiliation{Department of Physics, King's College London, Strand, London WC2R 2LS, UK}
\author{D. L. Pashov}
\altaffiliation[Present address: ]{ Department of Physics, King's 
  College London, Strand, London WC2R 2LS, UK}
\affiliation{Atomistic Simulation Centre, School of Mathematics
  and Physics, Queen's University Belfast, Belfast BT7 1NN, UK}

\begin{abstract}
  We present calculations of free energy barriers and diffusivities as
  functions of temperature for the diffusion of hydrogen in
  $\alpha$-Fe. This is a fully quantum mechanical approach since the
  total energy landscape is computed using a new self consistent,
  transferable tight binding model for interstitial impurities in
  magnetic iron. Also the hydrogen nucleus is treated quantum
  mechanically and we compare here two approaches in the literature both
  based in the Feynman path integral formulation of statistical
  mechanics. We find that the quantum transition state theory which
  admits greater freedom for the proton to explore phase space
  gives result in better agreement with experiment than the
  alternative which is based on fixed centroid calculations of the
  free energy barrier. We also find results in better agreement
  compared to recent centroid molecular dynamics (CMD) calculations of
  the diffusivity which employed a {\it classical\/} interatomic
  potential rather than our {\it quantum mechanical} tight binding
  theory. In particular we find first that quantum effects persist to
  higher temperatures than previously thought, and conversely
  that the {\it low temperature} diffusivity is smaller than predicted
  in CMD calculations and larger than predicted by classical
  transition state theory.  This will have impact on future modeling
  and simulation of hydrogen trapping and diffusion.
\end{abstract}


\maketitle

\section{Introduction}

The damaging effect of hydrogen (H) on the mechanical properties of
metals and alloys has been studied extensively since the
1940s. However, there is still considerable debate regarding the
specific mechanisms for H-assisted damage, that is, the role of H in
the fracture process.  Various doubts exist regarding the validity of
each of the existing H-assisted embrittlement mechanisms.\cite{{bi},
  {ka}} It is widely accepted that combinations of several mechanisms
can occur simultaneously and synergistically, since some involve
common background processes. A common factor in all H-assisted damage
mechanisms is the crucial role of H transport and trapping.  In all
cases, the predominant H-assisted damage mechanisms are dependent on
the rate and mode of H transport.

In particular, H diffusion in Fe and Fe alloys is extremely
important because it leads to engineering problems caused by H
embrittlement and degradation of high-strength steels.  Hydrogen in
$\alpha$-Fe diffuses between tetrahedral sites of the perfect bcc
lattice, the diffusivity being among the highest reported for any
metal.\cite{{km}, {hh1}, {hh2}, {nh}} This high H diffusivity
results from the very low activation energies due to the quantum
nature of H.\cite{ka}  The existence of microstructural imperfections
(vacancies, solute atoms, dislocations, grain boundaries, etc.)
introduces low energy trapping sites within the lattice which retard the
overall diffusion rate.\cite{{kh1},{kh2},{kj}} Because H is a light
element, intrinsic processes in H diffusion are strongly
influenced by its quantum mechanical behavior. At low temperatures
quantum tunneling is expected to be the dominant mechanism. At high
temperatures, the transition is dominated by classical jumping over
the barrier. In order to understand the process of H diffusion in Fe
it is essential to study H trapping and migration over the whole range
of temperatures covering both the quantum and classical dominated
regimes and the cross over between them.

The structure of the paper is as follows. In section~\ref{sec_MD} we
introduce the classical methodologies which we modify in
section~\ref{sec_PI} for the quantum nature of the diffusing
particle. We show the results of our calculations in
section~\ref{sec_results} in which we have combined the quantum
transition state theory\cite{v5} (QTST) for the first time with the
method of Wang and Landau (WLMC) for the calculation of free
energies.\cite{wl} This has the particular benefit that from a single
Monte-Carlo simulation the free energy may be extracted at any
temperature, in contrast to the usual Metropolis Monte-Carlo.\cite{tu}
Our results are demonstrated to be in excellent agreement with
experiment, and since our combined QTST WLMC scheme is very
compuationally efficient this opens the way to large scale simulations
of trapping bt defects in steel. We make some concluding remarks in
section~\ref{sec_conclusions}.

\section{Molecular dynamics and transition state theory}
\label{sec_MD}

One of the most commonly adopted approximations in atomistic
simulations of a system's evolution is the assumption that atomic
nuclei behave as classical particles.  Theoretical approaches based on
molecular dynamics (MD) have been used to study H trapping and
migration in Fe.\cite{{tm1}, {tm2}} An estimate of hydrogen
diffusivity can also be provided by employing the kinetic Monte Carlo
(kMC) method.\cite{{kmc1},{r}} The fundamental transition rate
constants used by kMC can be estimated without knowledge of the
dynamics of the system within the framework of the classical
transition state theory (TST).\cite{{r}, {r1}}  Unfortunately, when
simulations of H diffusion are made at or below room temperature
significant deviations from classical behavior are to be expected due
to the quantum nature of the proton motion.  An explicit treatment of
quantum effects is not only desirable for improvement of the accuracy
of the simulations, but it can be essential for understanding
phenomena and experimental observations depending directly on the
quantum nature of the nuclear motion.

The state of the art for quantum treatment of the ionic degrees of
freedom involves the use of the centroid path integral molecular
dynamics (CMD) method.\cite{jv} This was used recently to evaluate the
differences between the free energies of H at the interstitial and
binding sites in $\alpha$-Fe.\cite{{k1},{k2}} However, including
quantum effects is computationally demanding compared to a simulation
with classical nuclei, since one has to compute the energy of many
replicas of the physical system.  Studying H migration and
calculations of the diffusivity in the presence of microstructural
imperfections also require simulations in large blocks of atoms for
times exceeding the typical CMD time scales.

The kMC method has the advantage of being computationally less
expensive because the interatomic interactions are not computed
directly from the electronic structure “on the fly” as the simulation
proceeds.  Instead, kMC uses precomputed transition rates along the
minimal energy paths (MEP) between the metastable sites, thereby
allowing employment of more precise electronic structure methods,
density functional theory (DFT), tight binding (TB) or bond order
potentials for computation the energy of the physical system and
derivation of the transition rates.  The application of kMC for the
study of H diffusion permits simulations in larger blocks of atoms for
periods of time significantly longer than one can achieve with direct
MD simulation, which is essential for studying H migration and
trapping in the presence of microstructural imperfections and
consequent extraction of the diffusion coefficients.

Classical TST assumes that the H transition rates between metastable
sites follow approximately the Arrhenius law and the
activation energy is the difference of the energy for a fully relaxed
system at the saddle point separating the stable sites and the stable
site itself.\cite{{r}, {r1}}  Modern {\it ab initio} modeling
provides a good description of the energetics and potential energy
surface (PES) in Fe--H systems. However, these calculations of energy
barriers can not account for quantum corrections arising from the
low mass H atom.  Although the overall energy barriers are small, as
expected for a small atom like H and the geometry of the bcc lattice,
they are significantly higher than the experimentally determined
activation energies.\cite{r}  A quantum treatment of the hydrogen
degrees of freedom is mandatory to capture such effects.

\section{Feynman path integrals and quantum transition state theory}
\label{sec_PI}

Gillan\cite{{g1},{g2}} has argued that the appropriate quantum
generalization of the activated rate constant can be obtained using
the Feynman path integral (PI) method.\cite{fh} This
generalization involves the ratio of probabilities for finding the
centroid of the quantum chain at the saddle point and at the stable
site.  The activation energy is the difference of the free energy for
a fully relaxed system with the the centroid at the saddle point
separating the stable sites, and at the stable site itself.  Gillan has
also proposed a technique for calculation of this ratio in
path integral simulation.  Although Gillan's approach allows one to
examine the relative transition rates at different temperatures it
does not yield an absolute value for the activated rate constant, A
general TST like theory for calculation of the quantum activated rate
constant providing expressions for both the quantum activation free
energy and the prefactor was proposed by Voth
\cite{{v2},{v3},{v4},{v5}} Voth's quantum transition state theory
(QTST) presents the general quantum transition rate problem from the
perspective of path integral centroid statistics.\cite{fh}

In this paper, we study the activated dynamics of hydrogen
diffusion between tetrahedral sites in $\alpha$-iron by employing
Voth's path integral formulation of quantum transition state theory
for calculation of corresponding activation rate constants.  Apart
from allowing simulations in larger blocks of atoms for periods of
time significantly longer than can be achieved with CMD simulation,
the advantage of employing kMC with a transition rate determined by PI
QTST for studying H diffusion is that one can use an accurate
electronic structure approach for computation of the energy of the
physical system and derivation of the quantum transition rates.

\subsection{Interatomic forces and total energy}

Recent work studying diffusion of interstitial H in $\alpha$-Fe by
applying the CMD approach\cite{{k1},{k2}} employed an interatomic
potential for describing the electronic structure of the Fe--H system
within the embedded atom method (EAM) formalism.\cite{eam3} Although
there is a large number of existing classical
potentials,\cite{{eam1},{eam2},{eam3}} which are certainly useful,
they all suffer from a particular drawback in that the underlying
classical EAM type models require a huge number of parameters needing
to be fitted to a very large training set of data.  This and the
rather opaque functional form of the interatomic interactions in the
classical potentials mean that while they are able to model many
properties quantitatively they are at risk of failure once they are
transferred into situations for which they were not fitted. A well
known example of this is the failure of all but one of the many
classical potentials for $\alpha$-Fe to simulate correctly the core
structure of the screw dislocation; TB models do not suffer from this
problem.

The electronic structure and interatomic forces in magnetic iron, both
pure and containing hydrogen impurities, in the present calculations,
have been described using a non orthogonal self consistent
tight binding model.\cite{{ap2}, {ap1}} The transferability of
the model has been tested against known properties in many
cases. Agreement with both observations and DFT calculations is
remarkably good, opening up the way to quantum mechanical atomistic
simulation of the effects of hydrogen in iron.\cite{ap1}  By
contrast with EAM potentials, the TB model used in this paper comprises a
correct quantum mechanical description of both magnetism and the
metallic and covalent bond.

\subsection{Theory}

The approach to a quantum mechanical TST, proposed by Voth, is based
on Feynman's formulation of quantum statistical mechanics.\cite{fh} in
which the partition function $Z$ of a particle moving in one dimension
having Hamiltonian
\begin{equation*}
 H=\frac{p^2}{2m}+V(x)
\end{equation*} 
and in equilibrium with a heat bath at inverse temperature $\beta=1/k_BT$
is written approximately using a discretization of the imaginary time Feynman
path integral as\cite{GillanNATO}
\begin{equation}
 Z \approx Z_P = \left(\frac{mP}{2 \pi\beta\hbar^2}\right)^\frac{P}{2} 
\int dx_1...dx_P \,\,\exp \left\{ -\beta\sum_{s=1}^P \left[
    \frac{1}{2}\frac{mP}{\hbar^2\beta^2}
(x_{s+1}-x_s)^2+P^{-1} V(x_s) \right] \right\}
\label{PI}
\end{equation} 
This expresses the remarkable mapping of the partition function of a
{\it quantum mechanical} particle onto that of a necklace of $P$ beads
connected by {\it classical} harmonic springs of stiffness
$mP/\hbar^2\beta^2$, each bead feeling in addition a potential energy
$V(r)/P$, for large $P$ much weaker than the true potental
energy. Note also that the {\it potential energy} of the springs in
the classical analog derives from the {\it kinetic energy} operator in
the Hamiltonian. The numerical estimate converges to the quantum limit
when the discretization parameter $P$ is chosen to be large enough. In
the $P\rightarrow\infty$ limit, as Feynman shows, the partition
function is written as a path integral
\begin{equation*}
Z = \int{\cal D}x(\tau)\,e^{-S/\hbar}
\end{equation*}
which is an integral over all closed paths $x(\tau)$ in configuration
space. The action integral is
\begin{equation*}
S[x(\tau)] = \int_0^{\hbar\beta}{\rm d}\tau
    \left[\frac{1}{2}m\dot x^2(\tau) + V(x(\tau))\right]
\end{equation*}

The key to obtaining the rate constant in QTST is to separate out
paths in the {\it multidimensional} coordinate space into a ``reaction
coordinate'' denoted $q(\tau)$ and the remaining coordinates
$\r(\tau)$.\cite{v5} $q(\tau)$ might be a path connecting two stable
configurations through a saddle point, or it may be a path constrained
to the dividing surface separating two stable configurations.\cite{vi}
One can then work with a reduced centroid density
\begin{equation*}
\rho_c(q_c, \textbf{r}_c) = 
\int {\cal D}q(\tau) {\cal D}\textbf{r}(\tau) \delta(q_c-q_0) \delta(\textbf{r}_c-\textbf{r}_0) \exp \{ -S[q(\tau),\textbf{r}(\tau)] / \hbar \}
\end{equation*} 
in terms of which a constrained partition function is
\begin{equation*}
 Z_c(q^*) = \int d \, \textbf{r}_c \, \rho(q^*, \textbf{r}_c)
\end{equation*} 
where $q^*$ is the transition state value of the reaction coordinate.  The
``centroid'' variable $q_0$ in path integration is defined along the $q$
direction by the expression
\begin{equation*}
 q_0 = \frac{1}{\hbar\beta}\int_0^{\hbar\beta}d\tau q(\tau)
\end{equation*} 
The transition rate constant $\kappa$ can then be expressed in
terms of $Z_c(q^*)$ as
\begin{equation*}
 \kappa^{\rm QTST} = \frac{\bar{v}}{2}\frac{Z_c(q^*)}{Z_R}
\end{equation*} 
where, $Z_R$ is the unconstrained ``reactant'' partition function for
the particle localized about a lattice or trap site, and $\bar{v}$ is
a velocity factor\cite{v5} which can be estimated by adopting a free
particle dynamical model along the $q$ direction and taking the
velocity from a Maxwell distribution.  In this way, $\bar{v}$ becomes
\begin{equation*}
 \bar{v}_{FP} = \left(\frac{2}{\pi m \beta}\right) ^{\frac{1}{2}}
\end{equation*} 
and the quantum transition state rate constant is
\begin{equation}
 \kappa^{\rm QTST} = \left(\frac{1}{2\pi m \beta}\right) ^{\frac{1}{2}}
 \frac{Z_c(q^*)}{Z_R}
\label{QTST1}
\end{equation} 
This expression is difficult to implement in practice because in any
numerical calculation via either molecular dynamics or Monte Carlo
that generates a canonical distribution we do not have direct access to
the partition function. It can be computed by generating a
canonical distribution if it can be expressed in terms of averages of
phase space functions. The QTST proton transfer rate constant may be
re-expressed as \cite{v4}
\begin{equation}
  \kappa^{\rm QTST} = \nu \exp(-\beta \Delta F_c)
\label{QTST2}
\end{equation} 
where $\nu$ is the frequency of oscillation of the proton in the $q$
direction and the difference between centroid free energies of
the reactant and transition state is
\begin{equation}
 \Delta F_c = -k_B T\, \ln\left(\frac{Z_c(q^*)}{Z_c(q_R)}\right)
\label{QTST3}
\end{equation} 
$q_R$ is the reactant state value of the reaction coordinate $q$.  The
free energy difference between the reactant and transition state can
be evaluated using one of the methods for determination of the free
energy profile along the reaction coordinate.  An immediate
disadvantage of the energy profile approaches is that it is necessary
to perform many simulations of a system at physically uninteresting
intermediate values. Only initial and final configurations correspond
to actual physical states, and ultimately we can only attach physical
meaning to the free energy difference between these two states.
Nevertheless, the intermediate averages must be accurately calculated
in order for the integration to yield a correct result.  The number of
physically uninteresting intermediate averages increases dramatically
when one tries to derive the temperature dependence of the quantum
transition rate constant. In this case, the free energy difference
between the reactant and transition state has to be evaluated for each
temperature under interest, because both path integral MD method
\cite{pr} and conventional Monte Carlo methods generate a
canonical distribution at a given temperature.

\subsection{Free energy calculation}

In this paper, we use the Wang-Landau Monte Carlo (WLMC) algorithm
\cite{wl} to calculate directly the partition functions participating
in the QTST activation factor.  The key idea of the WLMC method is to
calculate the density of states $\Omega (E)$ directly by a random walk
in energy space instead of performing a canonical simulation at a
fixed temperature. The WLMC approach allows one to estimate various
thermodynamic properties over a wide range of temperatures from a
single simulation run.  The canonical partition function can be
expressed in terms of the density of states $\Omega (E)$ as
\begin{equation}
 Z(\beta) = \int_0^\infty dE\, e^{-\beta E}\, \Omega (E)
\label{WLMC}
\end{equation} 
The approach of Wang and Landau is to sample the density of states
directly and, once known, calculate the partition function via
(\ref{WLMC}). 

The path integral form of the quantum partition function $Z_P(\beta)$
in the $P$-bead approximation (\ref{PI}) can be expressed as \cite{vl}
\begin{equation*}
 Z_P(\beta) = \left(\frac{mP}{2 \pi\beta\hbar^2}\right)^\frac{P}{2}
\int d\textbf{r} \exp \left(- \frac{S_1(\textbf{r})}{\beta} - \beta S_2(\textbf{r})\right)
\end{equation*} 
which reveals that the quantum partition function depends on two
temperature independent functions
\begin{align*}
 S_1(\r) &= \sum_{s=1}^P \frac{mP}{2\hbar^2}(\r_{s+1}-\r_s)^2 \\
 S_2(\r) &= \sum_{s=1}^P( P^{-1} V(\r_s) )                   \\
\end{align*} 
$S_1 (\r)$ is related to the kinetic energy of the system and
$S_2 (\r)$ accounts for the potential energy.  Hence the
density of states, which is an analogue of the classical density of
states function, would depend on two variables, $s_1$ and $s_2$
\cite{vl}
\begin{equation}
 \Omega(s_1, s_2) = \int {\rm d}\r\,\delta(s_1-S_1(\r))\,\delta(s_2-S_2(\r))
 \label{Omega}
\end{equation} 
 Then we can express $Z$ as 
\begin{equation*}
 Z(\beta) = \int_0^\infty ds_1 ds_2\, \exp(-\frac{s_1}{\beta}-\beta s_2)\, \Omega (s_1, s_2)
\end{equation*} 
With this density of states the partition function that describes the
thermodynamics of the quantum system can be calculated for any
temperature.  However, we can determine the density of states only up
to a multiplicative constant, $\Omega_0$, since this will not change the relative
measures at different energy levels.  The uncertainty in the density
of states leads to a multiplicative uncertainty in the quantum
transition rate~(\ref{QTST1}).  The quantum transition rate at a given
temperature can be derived independently from~(\ref{QTST2}) by using a
method for determination of free energy difference between the
transition and reactant states.  If one knows $\kappa^{\rm QTST}$ at a
given temperature, the multiplicative constant can be determined
from~(\ref{QTST1}) and~(\ref{QTST2}) by comparison of the
corresponding transition rates at the same temperature.  In the
present work we calculate the free energy difference appearing
in~(\ref{QTST2}) by using an extension of the Wang-Landau sampling
scheme to the problem of the free energy profile along reaction
coordinates.\cite{c} If the process of transition is monitored by a
switching variable $q$, the free energy profile is
\begin{equation}
 F(q) = - k_BT\,\ln P(q)
\label{FE}
\end{equation} 
where $P(q)$ is the probability density that the system is in a state
with reaction coordinate $q$. Since $\Omega(E)$ and $P(q)$ play
similar roles the Wang-Landau sampling scheme can be used to generate a
function that approaches the probability $P(q)$ over many Monte Carlo
passes with the following Metropolis acceptance rule:
\begin{equation*}
 \hbox{acc}(\textbf{r}_c^{\rm old}\rightarrow\textbf{r}_c^{\rm new}) = \hbox{min} \left[1,\frac{\rho_c^{\rm new}(q,\textbf{r}_c)}{\rho_c^{\rm old}(q,\textbf{r}_c)}
\frac{g(q^{\rm old})}{g(q^{\rm new})}\right]
\end{equation*} 
where $g$ is the Wang-Landau scaling parameter.\cite{c} The
multiplicative constant $\Omega_0$ can also be determined if the
quantum transition rate~(\ref{QTST1}) at high temperature (the classical
limit) is approximated by a transition rate determined by the
classical TST.

\section{Calculation of the hydrogen diffusivity in iron}
\label{sec_results}

\begin{figure*}
\caption{\label{Fig1} (color online)
Potential energy surfaces for a H~atom moving into the fixed atomic
positions of the Fe atoms; (a) configuration in which the
Fe atoms have been relaxed around a H~atom in the tetrahedral
``reactant'' site; (b) the Fe atoms are relaxed around a H~atom held at
the saddle point position between two neighboring tetrahedral sites.
        }
\begin{center}
\includegraphics[scale=0.5]{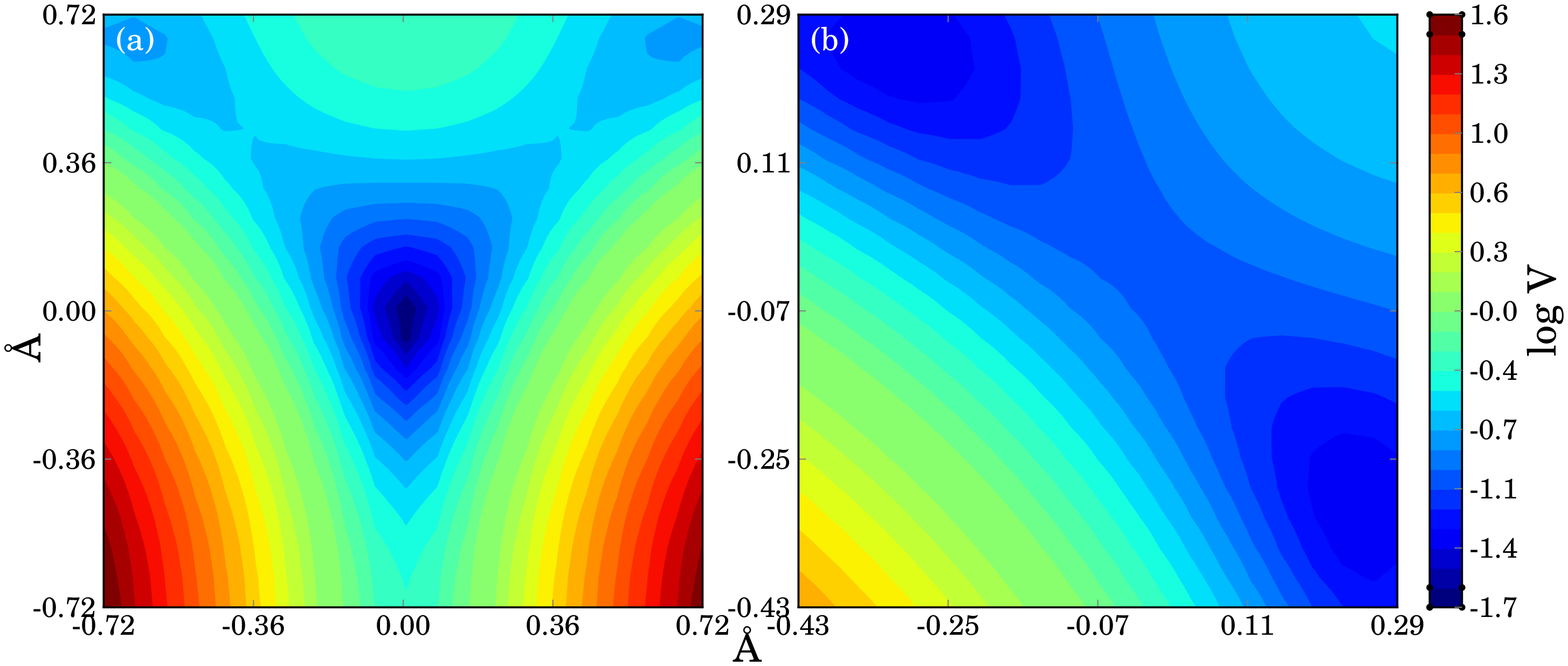}
\end{center}
\end{figure*}

Here we study the real time quantum dynamics of hydrogen diffusion in
perfect $\alpha$-iron by employing the path integral (PI) approach
described above combined with WLMC.  The electronic structure and
interatomic forces in magnetic iron, both pure and containing hydrogen
impurities, in the present calculations have been described using a
non orthogonal self consistent tight binding model.\cite{{ap1}, {ap2}}
It is to be noted that the TB model predicts, correctly, that the
configuration with the lowest potential energy is the tetrahedral
site.  The minimum energy path (MEP) and the transition state between
two adjacent tetrahedral sites have been identified by using the
nudged elastic band (NEB) method.\cite{neb} In order to use the
PI-WLMC technique to calculate corresponding partition functions, we
constructed 3D potential energy surfaces (PESs) for the hydrogen
motion in two fixed lattice configurations, corresponding to fully
relaxed lattice of 16 Fe atoms with H atom in tetrahedral and saddle
point configurations.  We constructed PESs in these fixed lattice
configurations by performing a large set of TB total energy
calculations corresponding to different positions of the hydrogen
nucleus.  In all these calculations the nuclei are treated as
classical point particles. The calculated potential energy surfaces
are shown in figure~\ref{Fig1}

\begin{figure*}
\caption{\label{Fig2} (color online)
 Density of states $\Omega(s_1, s_2)$ (\ref{Omega}) for the proton
 moving in the potential of fixed Fe atoms as shown in
 figure~\ref{Fig1}; (a)  centroid fixed at tetrahedral site;
(b) centroid in the vicinity of tetrahedral site; (c) centroid fixed
at the saddle point; (d) centroid on the dividing surface in the
region of the barrier top. (a) and (c) are those used in Gillan's
theory in which the centroids are fixed at reactant and saddle point
positions. (b) and (d) are appropriate to Voth's method; in (b) the
centroid is allowed to explore phase space in the region of the
tetrahedral lattice site and in (d) the centroid is confined to the
Vineyard dividing surface that intersects the saddle point.
        }
\begin{center}
\includegraphics[scale=0.5]{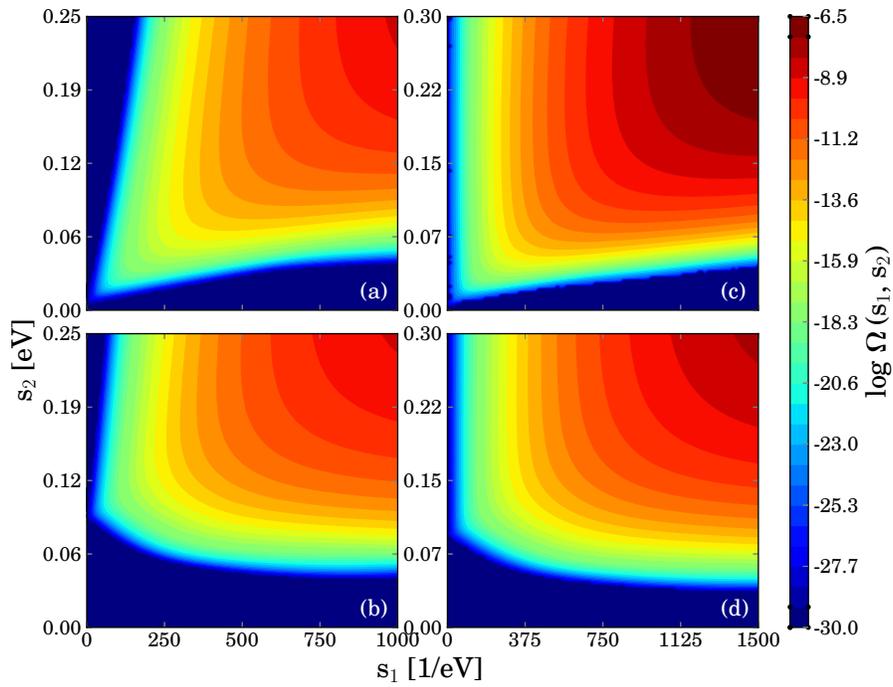}
\end{center}
\end{figure*}

We have studied the importance of quantum effects in hydrogen
diffusion in perfect bcc-Fe by employing both Gillan's approach for
calculation of the activation energy and Voth's formulation of path
integral QTST.  Our main result is the calculation of the density of
states~(\ref{Omega}) at the stable state and in the region of the
barrier top of a Fe--H system in the cases of Gillan's formulation of
the activation barrier and Voth's PI QTST generalization of transition
state theory. With these densities of states (DOS) the corresponding
partition functions, describing the thermodynamics of the quantum
system, can be calculated for any temperature. Knowing corresponding
partition functions, we can determine a number of thermodynamic
variables for the Fe--H system as functions of the temperature. The DOS of
the stable and transition states as they are defined in both Gillan's
approach and path integral QTST are shown in figure~\ref{Fig2}

\begin{figure*}
  \caption{\label{Fig3} (color online) Position probability density
    (PPD) of the hydrogen nucleus when the centroid is fixed at
    tetrahedral site at temperature of (a) $T=$20K; (b) $T=$50K; (c)
    $T=$100K; (d) $T=$200K; (e) $T=$ 300K; (f) $T=$1000K. In each
    panel the tetrahedral site is located at the center of the
    image. Note that at high temperature the PPD is spherical and
    located close to the centroid as expected of a classical
    particle. At intermediate lower temperatures the PPD spreads out,
    but at the lowest temperature the PPD contracts due to the
    freezing of the proton into its lowest oscillator state; however
    it is no longer spherical as it ``feels'' the low symmetry of the
    tetrahedral site.  }
\begin{center}
\includegraphics[scale=0.5]{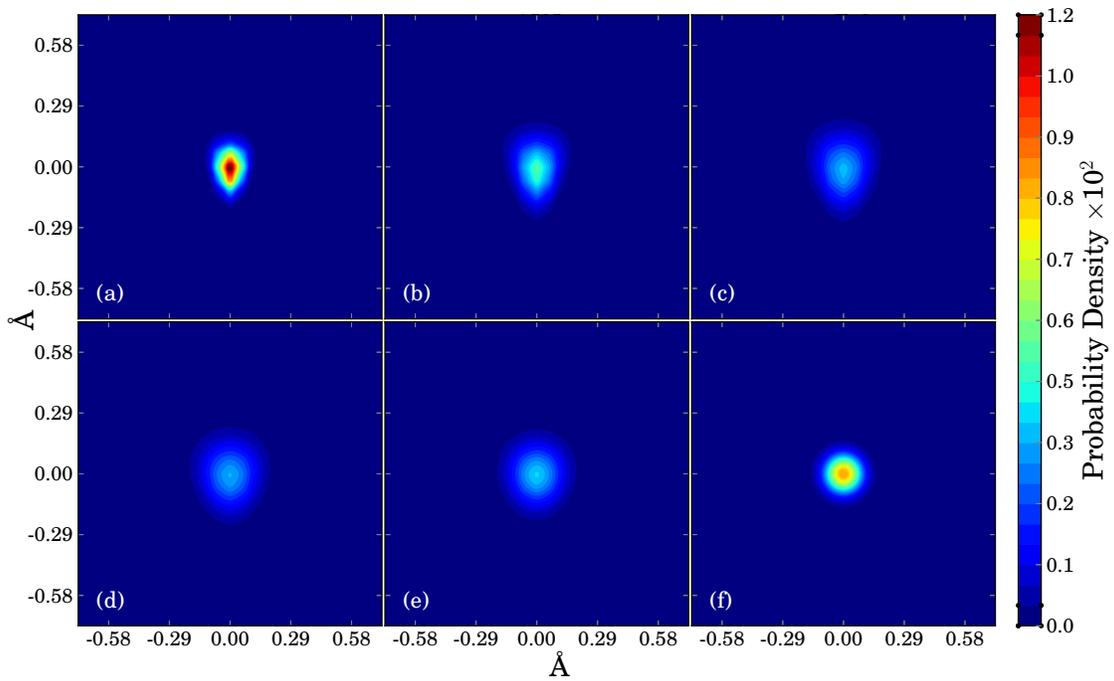}
\end{center}
\end{figure*}

\begin{figure*}
  \caption{\label{Fig4} (color online) Position probability density of
    the hydrogen nucleus when the centroid is fixed at the saddle
    pointe at temperature of (a) $T=$20K; (b) $T=$50K; (c) $T=$100K;
    (d) $T=$200K; (e) $T=$ 300K; (f) $T=$1000K. In each panel the
    saddle point is at the center of the image and there is a symmetry
    equivalent tetrahedral site at the upper left and lower right
    corners. Note how, at low temperature the proton ``splits into
    two'' and even though the centroid of the chain of ``beads'' is
    held fixed at the saddle point the greatest position probability
    density is close to the energy minima at the tetrahedral
    ``reactant'' sites.  }
\begin{center}
\includegraphics[scale=0.5]{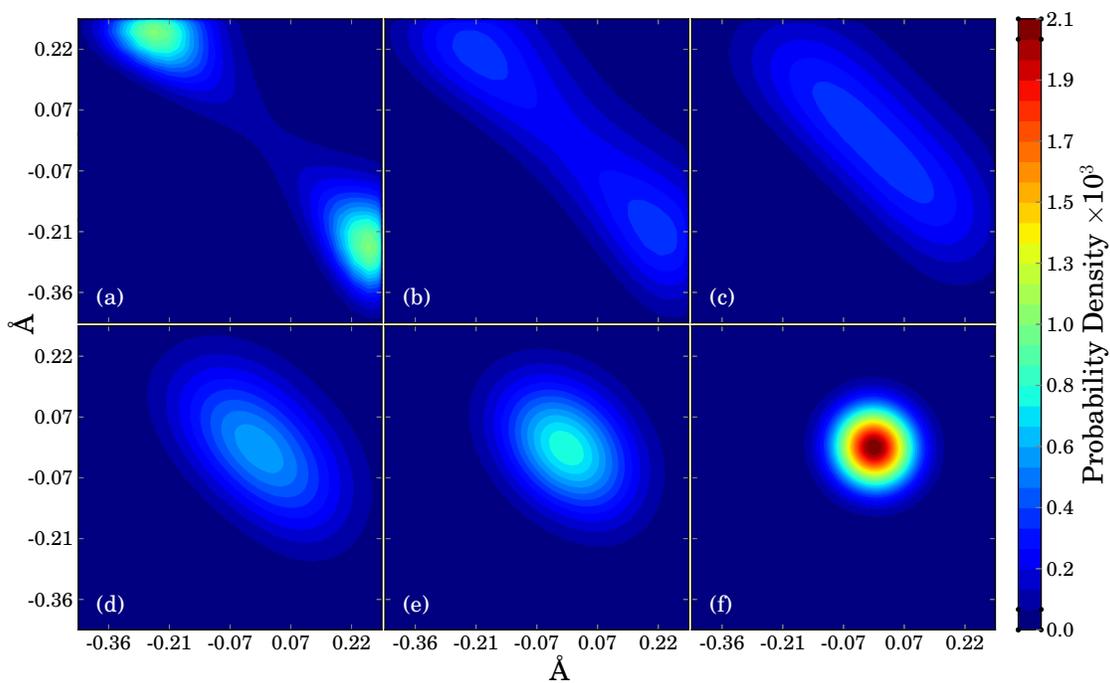}
\end{center}
\end{figure*}

\begin{figure*}
\caption{\label{Fig5} (color online)
        Position probability density of the hydrogen nucleus when the
        centroid is in the vicinity of tetrahedral site at temperature
        of (a) $T=$20K; (b) $T=$50K; (c) $T=$100K; (d) $T=$200K; (e)
        $T=$300K; (f) $T=$1000K.  In each panel the tetrahedral site
        is located at the center of the image. In comparison to
        figure~\ref{Fig3} the particle is more spread out even at the
        highest temperature. This reflects the greater degree of
        freedom of the PI QTST theory compared to the method of Gillan.}
\begin{center}
\includegraphics[scale=0.5]{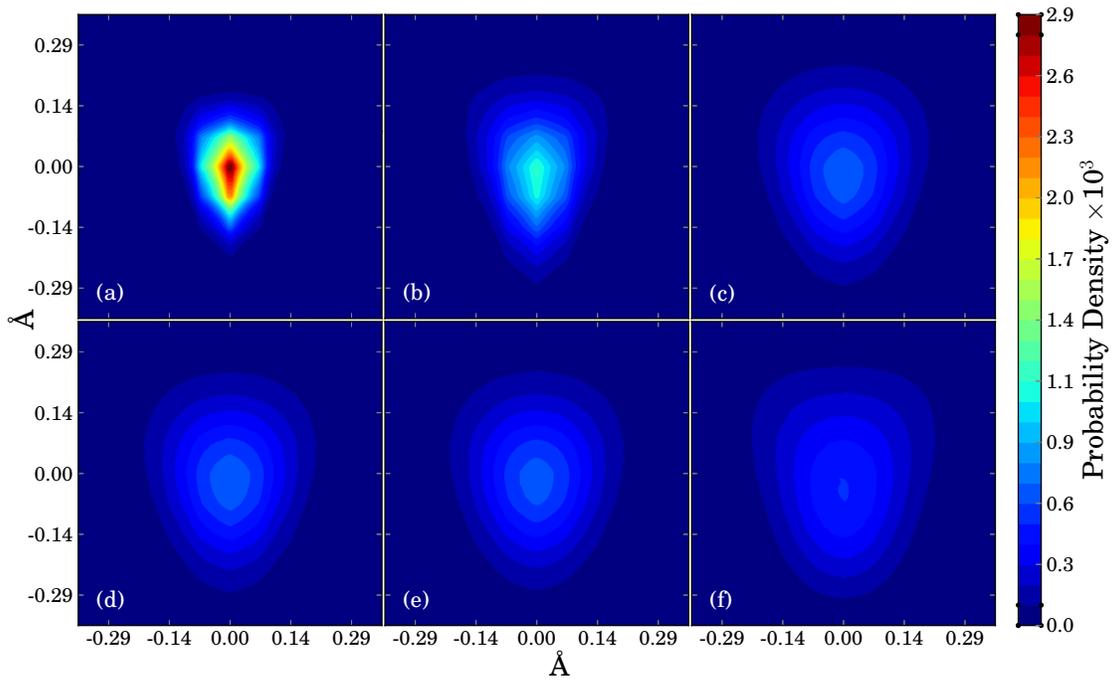}
\end{center}
\end{figure*}

\begin{figure*}
  \caption{\label{Fig6} (color online) Position probability density
    (PPD) of the hydrogen nucleus when the centroid is on the dividing
    surface in the region of the barrier top at temperature of (a)
    $T=$20K; (b) $T=$50K; (c) $T=$100K; (d) $T=$200K; (e) $T=$ 300K;
    (f) $T=$1000K. As remarked in the caption to figure~\ref{Fig5} the
    PPD is greatly spread out compared to that in the Gillan
    formulation of the path integral method. In addition, especially
    at high temperatures it is seen that the reduced density is
    allowing the proton to explore the phase space along the Vineyard
    dividing surface, which is ``perpendicular'' to the reaction
    path. As the centroid is no longer constrained to remain at the
    saddle point the resulting free energy barrier is larger and the
    diffusivity smaller in Voth's method. }
\begin{center}
\includegraphics[scale=0.5]{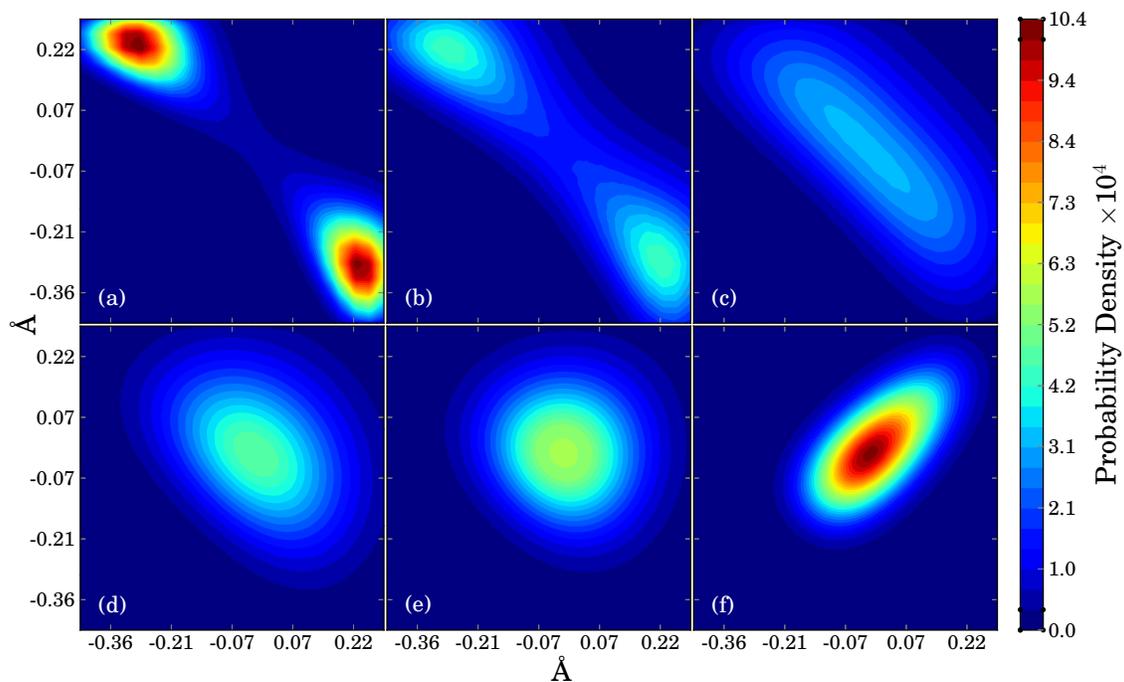}
\end{center}
\end{figure*}

We also determine the position probability density of the hydrogen
nucleus (proton) when it is in tetrahedral and saddle point
configurations.  In Gillan's path integral procedure H is in stable or
transition states when the centroid of the imaginary time reaction
coordinate path is in a stable interstitial impurity site
(figure~\ref{Fig3}) or a saddle point between two neighboring
interstitial sites (figure~\ref{Fig4}).  Within Voth's formulation of
QTST, we determine the position probability density of the hydrogen
nucleus when its centroid is confined in the potential well of
tetrahedral sites (figure~\ref{Fig5}) and on the dividing surface that
intersects the classical saddle point between two such sites
(figure~\ref{Fig6}).  The probability distributions as functions of
temperature, when the centroid is at the transition state, are shown
in figure~\ref{Fig4} and figure~\ref{Fig6}. It is very clear that in
both cases at low temperature the proton in the transition state
``splits into two'' with greatest position probability density not at
the saddle point but very close to the tetrahedral sites.  At low
temperatures the transition rate is dominated by quantum tunneling
through the barrier and both Gillan's and Voth's approaches predict
similar results.  In the classical limit (high temperature) the
hydrogen atom can be considered as a classical particle and its
position probability density is concentrated in the vicinity of the
potential well and classical saddle point. At high temperature, the
reduced centroid density $Z_c(q^*)$ and the reactant partition function
$Z_R$ of a Fe--H system in a stable tetrahedral site can be
approximated by the configurational partition functions of a classical
particle existing in quasi equilibrium.  The assumption that the
motion of atoms in both configurations can be treated as simple
harmonic oscillators leads to the familiar Vineyard--Slater
expression,\cite{vi} derived from many body transition state theory.
Hence, the position probability density of the hydrogen nucleus
calculated in the framework of path integral QTST (figures~\ref{Fig5}
and~\ref{Fig6}) includes corrections due to motion (thermal
``vibration'') of the proton near the stable state and in the region
of the barrier top.  As seen from the comparison between
figure~\ref{Fig3} and figure~\ref{Fig5}, as well as between
figure~\ref{Fig4} and figure~\ref{Fig6}, these corrections are
significant at room temperature and at high temperature.

\begin{figure*}
  \caption{\label{Fig7}  The free energy needed to carry
    the H atom from an initial stable position to a transition state
    as a function of temperature, obtained using Gillan's
    approach. The activation energy is compared with assessments from
    hydrogen equilibration and permeation tests in ref.~[\onlinecite{km}].
          }
\begin{center}
\includegraphics[scale=0.5]{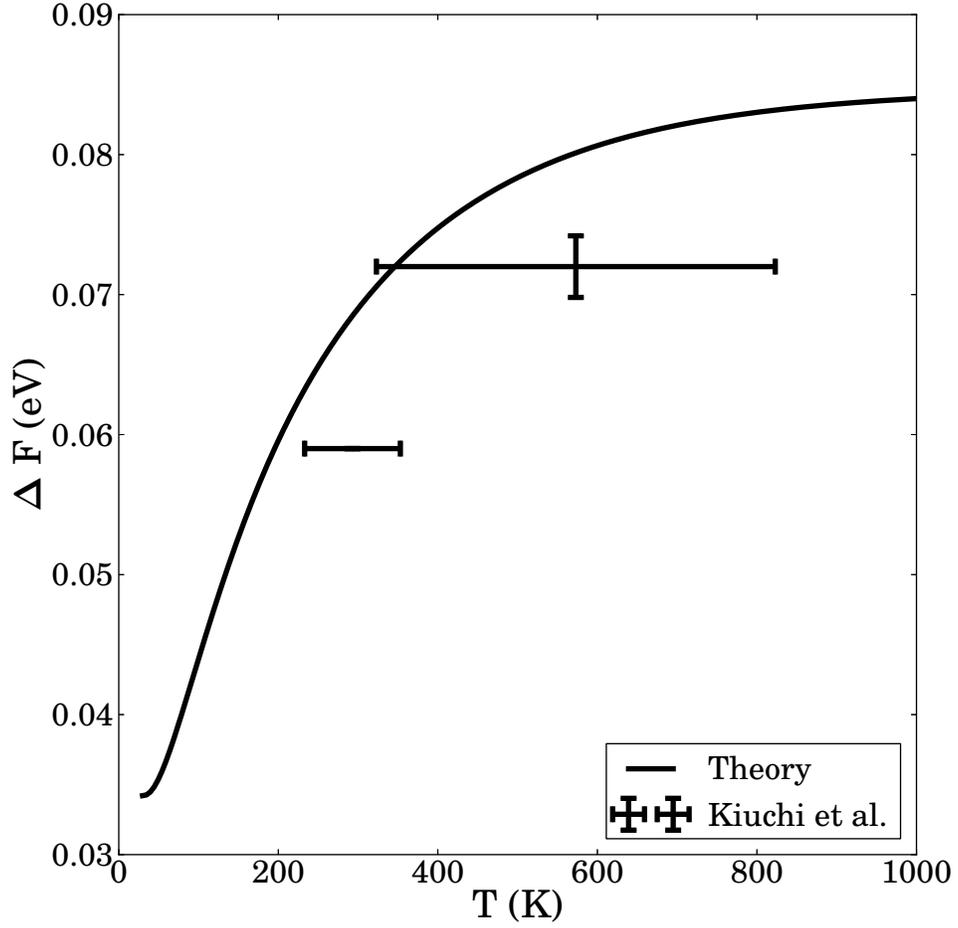}
\end{center}
\end{figure*}

In order to determine the activation barrier height given by Gillan's
approach and the PI QTST transition rate as a function of temperature
we have calculated the corresponding partition functions appearing in
both methods in the temperature interval between 20 and 1000K.  Since
the DOS are determined by a WLMC path integral approach up to a
multiplicative constant, we apply the extension of Wang-Landau
sampling to the problem of the free energy profile to obtain the
corresponding probability distribution functions $P(q)$ at a fixed
temperature of 1000K.  We find that Gillan's free energy difference
between stable and transition states is $\Delta F=0.084$eV at 1000K.
As is to be expected this value is very close to the classical limit
of the migration barrier ($E_m= 0.088$eV) because quantum corrections
are negligible at high temperatures.  After determination of the
corresponding multiplicative constant, $\Omega_0$, we can calculate
the free energy needed to carry the H~atom from an initial stable
position to a transition state in the temperature interval between 20
and 1000K.  The activation energy, defined by Gillan's quantum
generalization, as a function of temperature is shown in
figure~\ref{Fig7}.

\begin{figure*}
\caption{\label{Fig8} 
 Quantum transition rate calculated by PI QTST in the temperature interval 20--1000K.  
        }
\begin{center}
\includegraphics[scale=0.5]{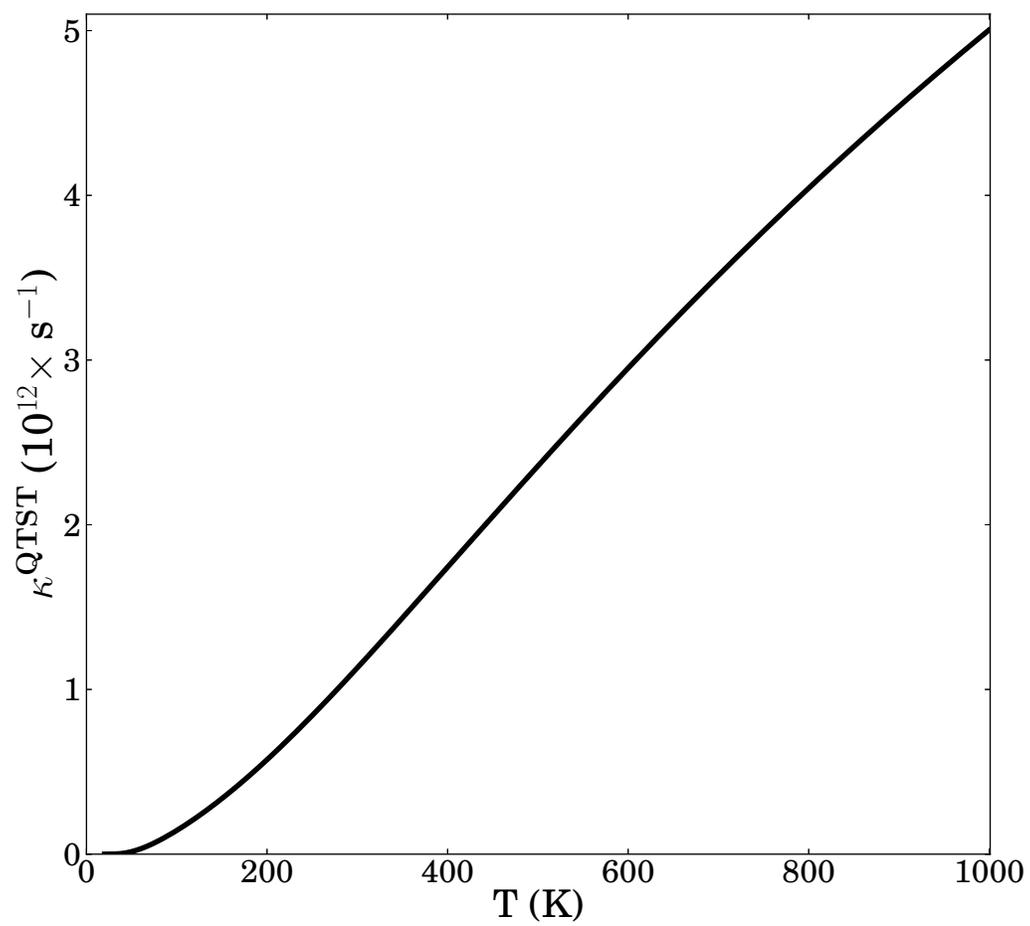}
\end{center}
\end{figure*}

The free energy difference between transition and reaction states at
1000K, given in Voth's theory by the ratio between the corresponding
reduced centroid densities (\ref{QTST3}), is calculated by using WLMC
to obtain the free energy profile (\ref{FE}). We find that the
free energy needed to carry a H~atom from a stable to a transition state
at 1000 K is 0.087eV.  The proton transition rate at 1000K is
determined from (\ref{QTST2}). The value of frequency of oscillation of the
proton in the $q$ direction, $ \nu = 1.29 \times 10^{13}$s$^{-1} $,
appearing in (\ref{QTST2}) is derived from a harmonic fit to the energy
along the reaction coordinate determined by NEB TB calculations.  After
calculation of the multiplicative constant $\Omega_0$ we can find the
temperature dependence of the quantum transition rate between two
tetrahedral sites in $\alpha$-Fe in the interval between 20 and 1000K
(see figure~\ref{Fig8}). This is the primary goal of the current work.

\begin{figure*}
  \caption{\label{Fig9} (color online) Diffusion coefficients of H in
    $\alpha$-Fe in the temperature range 100--1000 K calculated by PI
    QTST (blue line).  The pink band of data represents an assessment
    by Kiuchi and McLellan\cite{km} of hydrogen gas equilibration
    experiments. The solid red and green lines are data from
    electrochemical permeation experiments, assessed in
    ref.~[\onlinecite{km}] and measured in ref.~[\onlinecite{nh}]
    respectively. The blue band of data are measurements by Grabke and
    Rieke\cite{gr} and the yellow line shows measurements by
    Hayashi~{\it et~al.}\cite{hh2} The triangles show theoretical results from
    centroid molecular dynamics calculations using a classical
    interatomic potential and are taken from
    ref.~[\onlinecite{k1}]. Note that these are in less good agreement
    with experiment at high temperature compared to our fully quantum
    mechanical predictions and that at low temperatures the CMD method
    predicts diffusivities significantly larger than ours. This may
    reflect the difficulty in Metropolis Monte Carlo sampling at low
    temperature in contrast to the WLMC method which accesses all
    temperatures as a result of a single sampling.}
\begin{center}
\includegraphics[scale=0.5]{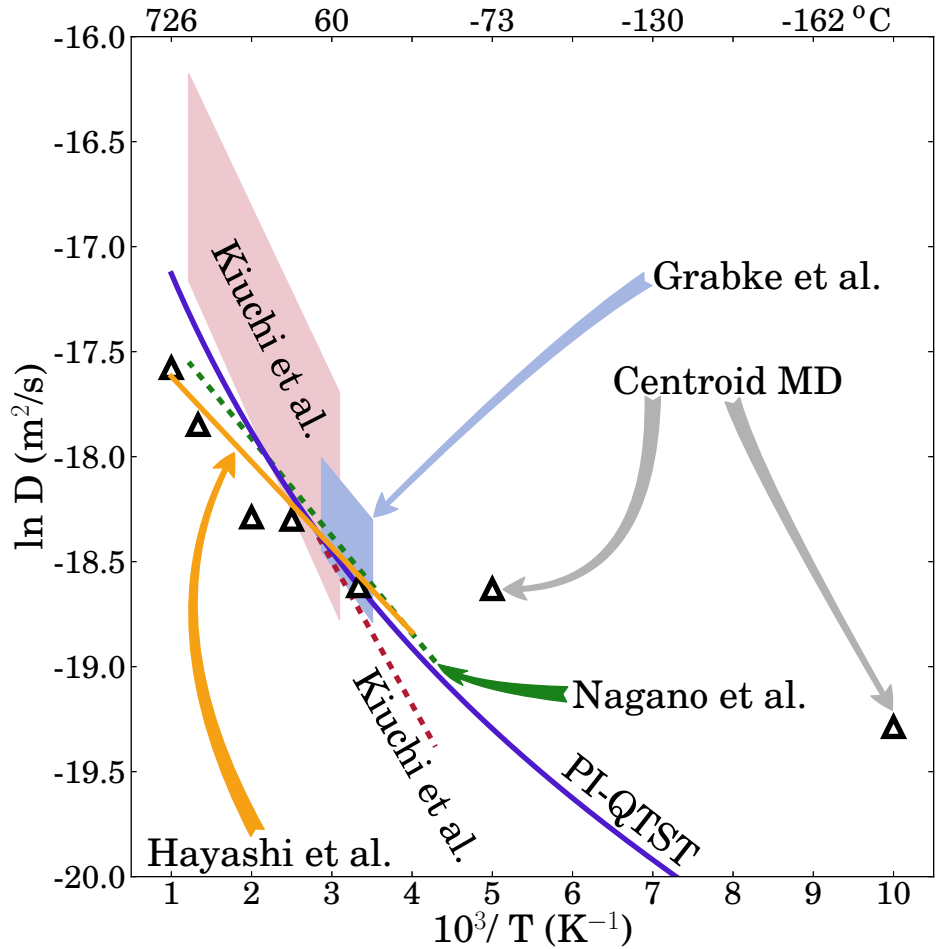}
\end{center}
\end{figure*}

In the simple case of H diffusion in a perfect bcc lattice, the
diffusion coefficient $D$ can be determined from the Einstein formula
assuming an uncorrelated random walk in cubic symmetry
\begin{equation*}
D = \frac{1}{6} z  R^2 \kappa^{\rm QTST}
\end{equation*} 
where $z=4$ is the number of neighboring positions and $R=a^2/8$ is
the jump distance. An Arrhenius plot of the diffusion coefficient as a
function of temperature in the interval 20 to 1000K is shown in
figure~\ref{Fig9}. Diffusion coefficients determined at different
temperatures by kMC using the same $k^{\rm QTST}$ confirm the results
obtained from the Einstein formula.  For comparison, experimental
diffusivities over a wide temperature range
(240–-1000K)\cite{nh,km,hh1,hh2} and diffusion coefficients calculated
by the centroid molecular dynamics (CMD) method\cite{{k1}, {jv}} at
several temperatures between 100 and 1000K are also plotted in
figure~\ref{Fig9}. Our results are in very reasonable agreement with
the experimental measurements.  A deviation from linear behavior is
observed in the Arrhenius plots based on our PI QTST results.  Also,
they are in good agreement with the diffusion coefficients calculated
by using computationally demanding centroid molecular dynamics
technique\cite{k1} in the interval 300--1000K, while at low
temperatures these data show a much larger diffusivity. It should be
noted that in view of the logarithmic axis in figure~\ref{Fig9} our
quantum mechanical tight binding predictions are in better agreement
even at high temperature with those using an energy landscape based in
a classical interatomic potential. Our PI QTST results are
in excellent agreement with the experimental measurements below 300 K,
where CMD predicts larger diffusion coefficients.

\section{Concluding remarks}
\label{sec_conclusions}

Our kMC path integral QTST approach in combination with WLMC, along
with the TB model describing energy of the Fe--H system, permits a
description of real time quantum dynamics of hydrogen diffusion in
$\alpha$-iron over a wide temperature range. Unlike conventional MD
and Monte Carlo methods generating canonical distributions, and
computationally demanding centroid MD techniques, the WLMC algorithm
allows one directly to calculate the partition functions participating
in the PI QTST transition rates over a wide range of temperatures from
a single simulation run.  The results reveal that quantum effects play
a crucial role in the process of H migration even at room
temperature.  Although Gillan's quantum generalization of the
activated rate constant describes correctly quantum tunneling at low
temperatures, the thermal ``vibrations'' of the proton at the saddle
point are not accounted for within this approach, which leads to an
underestimation of the activation barrier height at room and high
temperatures.  The diffusion coefficient as a function of temperature,
calculated using the QTST rate constant determined by PI
WLMC, is in good agreement with the experimentally evaluated
diffusivity in the interval 240–-1000K.

The computationally less expensive kMC method using precomputed PI
QTST transition rates calculated by WLMC technique, proposed in this
article, opens the way to studying the quantum dynamics of hydrogen
migration and trapping in presence of microstructural imperfections
and calculation of the diffusion coefficients over a wide temperature
range.

\section*{Acknowledgments}

We are grateful to the European Commission for funding under the
Seventh Framework Programme, grant number 263335, MultiHy (multiscale
modelling of hydrogen embrittlement in crystalline materials). We are
glad to acknowledge discussions with Nicholas Winzer and Matous
Mrovec.

\def\JPCM{J.~Phys.:~Condens.~Matter.}
\def\PRB{Phys.~Rev.~B}
\def\PRL{Phys.~Rev.~Lett.}

%



\end{document}